\documentclass[aps,physrev,twocolumn,superscriptaddress, longbibliography]{revtex4-2}  
\usepackage{graphicx}  
\usepackage{dcolumn}   
\usepackage{bm}        
\usepackage{amssymb}   
\usepackage[utf8]{inputenc}  
\usepackage[T1]{fontenc}

\usepackage{xcolor}
\usepackage{parskip}

\usepackage{url}

\usepackage{hyperref}
\newcommand{\equalcontribution}{These authors contributed equally to this work.}

\begin{document}

\title{Magnetic landscape of NbTiN superconducting resonators under radio-frequency excitation}

\author{J. Baumgarten}
\thanks{\equalcontribution}
\affiliation    {Experimental Physics of Nanostructured Materials, Q-MAT, Université de Liège, Allée du 6 Août 19, Sart Tilman B-4000, Belgium.}

\author{N. Lejeune}
\thanks{\equalcontribution}
\affiliation    {Experimental Physics of Nanostructured Materials, Q-MAT, Université de Liège, Allée du 6 Août 19, Sart Tilman B-4000, Belgium.}

\author{L. Nulens}
\affiliation{Quantum Solid-State Physics, Department of Physics and Astronomy, KU Leuven, Celestijnenlaan 200D, Leuven, B-3001, Belgium}

\author{I. P. C. Cools}
\affiliation{Department of Microtechnology and Nanoscience, Chalmers University of Technology, Gothenburg, SE-412 96, Sweden}


\author{J. Van de Vondel}
\affiliation{Quantum Solid-State Physics, Department of Physics and Astronomy, KU Leuven, Celestijnenlaan 200D, Leuven, B-3001, Belgium}

\author{A. V. Silhanek}
\affiliation    {Experimental Physics of Nanostructured Materials, Q-MAT, Université de Liège, Allée du 6 Août 19, Sart Tilman B-4000, Belgium.}

\date{\today}

\begin{abstract}
Planar superconducting resonators are essential components in quantum circuits and highly sensitive sensors. However, their performance is often compromised by magnetic flux penetration, as the interaction of flux quanta and the induced radio-frequency (RF) currents in the superconducting thin film leads to significant energy dissipation. At low operating temperatures, this issue is aggravated as thermomagnetic instabilities can trigger the sudden propagation of magnetic flux avalanches. An important open question is whether the RF excitation itself stimulates the nucleation and propagation of magnetic flux avalanches in the superconducting thin film. The literature remains inconclusive on this point, partly due to the lack of compelling evidence for this phenomenon. In this work, we address this issue by unprecedented direct visualization of magnetic flux penetration through Faraday rotation imaging under simultaneous RF excitation. We demonstrate that the avalanche activity exhibits a weak dependence on the RF intensity for RF excitations within the linear Campbell regime. However, magnetic flux bursts clearly influence the RF transmission properties of the device. Furthermore, it is possible to unambiguously associate a particular avalanche event with a jump in resonance frequency. This enables us to identify the loci of most deleterious events and understand the distinct origins of upward and downward frequency shifts. These observations are supported by electromagnetic simulations in which local changes of the kinetic inductance mimic flux avalanches and confirm the invasive character of the MOI technique. The insights gained from this study aim to contribute to the broader understanding of the magnetic resilience of superconducting resonators, with the goal of improving their efficiency and stability.
\end{abstract}
\maketitle

\section{Introduction}
Planar superconducting resonators operating at radio frequencies have driven major advances in a wide range of applications, including quantum sensing and detection \cite{day_broadband_2003,Tominaga2025}, quantum computing \cite{wallraff_strong_2004,Mahashabde-2020}, and information processing \cite{blais_quantum_2020,Joshi_2021}. Their unique advantage stems from the superconducting properties of the materials composing these resonators, notably their zero electrical resistance. The aforementioned applications require low energy losses in the system, characterized by high quality factors (Q), which is the ratio between the average stored magnetic energy and the dissipated energy. Indeed, superconducting resonators have been known to reach quality factors up to nearly 10$^{8}$, several orders of magnitude higher than their metal counterparts \cite{Tominaga2025}.

To achieve such performances, it is crucial that the superconducting resonator remains free of superconducting vortices, as their motion leads to significant energy dissipation and consequently deteriorates the quality factor \cite{song_microwave_2009}. Although magnetic screening methods can be used to mitigate this problem, some applications require exposure to an external magnetic field \cite{wilde2025}, such as certain qubit implementation schemes like spin ensembles in solid-state systems \cite{kubo_strong_2010}, phase-slip qubits \cite{astafiev_coherent_2012}, trapped electrons \cite{bushev_trapped_2011}, and many others \cite{uilhoorn2021,Wang2023,elhomsy2023,janssen2025}. In addition, even though planar superconducting resonators are resilient to in-plane magnetic fields \cite{kim2025}, unwanted small in-plane misalignments can give rise to out-of-plane magnetic field components that can severely affect their performance. 

Generally, minimizing vortex-induced losses implies pinning them in order to prevent their motion. Another approach is to decrease the number of vortices in regions of high RF current amplitude and/or magnetic fields. Multiple studies explored the first approach, by introducing artificial pinning sites \cite{bothner_improving_2011, song_reducing_2009, nsanzineza_trapping_2014}.  
An alternative method involves modifying the ground plane, with studies showing that reducing its size \cite{bothner_improving_2017, cools_losses_2025} or segmenting it into fractal structures \cite{graaf_magnetic_2012} can effectively alleviate demagnetization effects (flux focusing) and minimize field penetration.
While the strategies based on increasing vortex pinning can help reduce vortex-induced losses at temperatures not too far from the superconducting transition $T_c$, at temperatures below $\sim T_c/2$, they may favor localized thermomagnetic instabilities~\cite{menghini_dendritic_2005}. These instabilities can trigger flux avalanches, which involve the rapid nucleation and propagation of magnetic flux into the superconducting material \cite{vestgarden_lightning_2012}, leading to performance degradation in superconducting resonators. Indeed, these flux avalanches have a significant impact on the resonators, introducing stochastic and unpredictable behavior that manifests as fluctuations in the resonance frequency as a function of the applied magnetic field \cite{nulens_catastrophic_2023}. Technologically relevant resonators based on conventional superconductors are particularly susceptible to these events, whereas flux avalanches are seldom detected in hight temperature superconductors \cite{Velluire-Pellat2023,roitman_characterization_2022,Lejeune-2025}.

Surprisingly, despite extensive research on the properties and behavior of superconducting resonators, the specific role of RF excitation in nucleating flux avalanches remains largely unexplored. There have been experimental studies suggesting that RF excitation does promote flux
avalanches \cite{ghigo_evidence_2007,ghigo-2009,awad_flux_2011,Cuadra-Solís2011}, while others report a suppression of magnetic flux avalanches for probing frequencies nearing the depinning frequencies \cite{lara_thermally_2017}. From a theoretical standpoint, the information on the matter is also scarce. Numerical simulations have been used to investigate the effect of a uniform AC magnetic field of relatively low frequency (a few MHz) and show that the AC excitation reduces the threshold field for the onset of flux avalanches \cite{jiang_sensitivity_2022}.  It should also be pointed out that experimental evidence \cite{Yogi} and theoretical predictions \cite{Gurevich_2023} indicate that at temperatures close to the critical one, the first-vortex penetration field for RF magnetic fields is somewhat larger than for DC fields. This could be beneficial to keep vortex penetration at bay. However, it is expected that this difference becomes less pronounced at low temperatures (see \cite{Sheikhzada-2020,Wang-2022} and references therein).

The aim of this work is to provide further insights into the correlation and possible causation relation between RF excitations and magnetic flux avalanches in superconducting resonators. To this end, radio-frequency transmission measurements are combined with widefield magneto-optical imaging, allowing, on the one hand, for simultaneous and direct association of the nucleation and propagation of flux avalanches and their impact on the resonance frequency. On the other hand, we investigate if the presence of localized RF excitation influences the statistical distribution of the size and occurrence rate of magnetic flux avalanches. We demonstrate that the required concomitant radio-frequency measurements and magneto-optical imaging are a challenging feat, as the presence of the magnetic indicator significantly impacts the response of the resonator. We establish a clear correspondence between individual avalanche events and discrete shifts in resonance frequency, enabling the spatial mapping of the most deleterious occurrences and clarifying the distinct origins of upward and downward frequency shifts. Electromagnetic simulations, in which localized variations of the kinetic inductance emulate flux avalanches, reproduce these effects and help to underscore the invasive nature of the MOI technique. Together, these results establish a direct link between magnetic flux avalanches and resonance behavior in superconducting resonators, highlight the location of the most deleterious events, and inform future approaches for enhancing their stability and performance.

\section{Experimental Methods}

\subsection{Sample fabrication}
The resonators used in this work were fabricated either on double-side polished C-plane $\mathrm{Al_2O_3}$ wafers of thickness 430 µm or on 525 µm-thick undoped Si. A 100 nm thick NbTiN film was deposited by sputtering from a $\mathrm{Nb_{81.9}Ti_{18.1}N}$ target in a near UHV environment, with in-situ annealing at 660°C. The coplanar waveguide resonators were patterned using electron beam lithography (PMMA A4 resist with an E-spacer conducting layer) followed by Ar/Cl inductively coupled plasma etching. Residual resist was removed by heated acetone. The fabrication was carried out at MyFab, Chalmers University of Technology. A more detailed description of the fabrication process of these samples can be found in Ref. \cite{nulens_catastrophic_2023}.

\begin{figure*}[!ht]
\centering
\includegraphics*[width=1\linewidth]{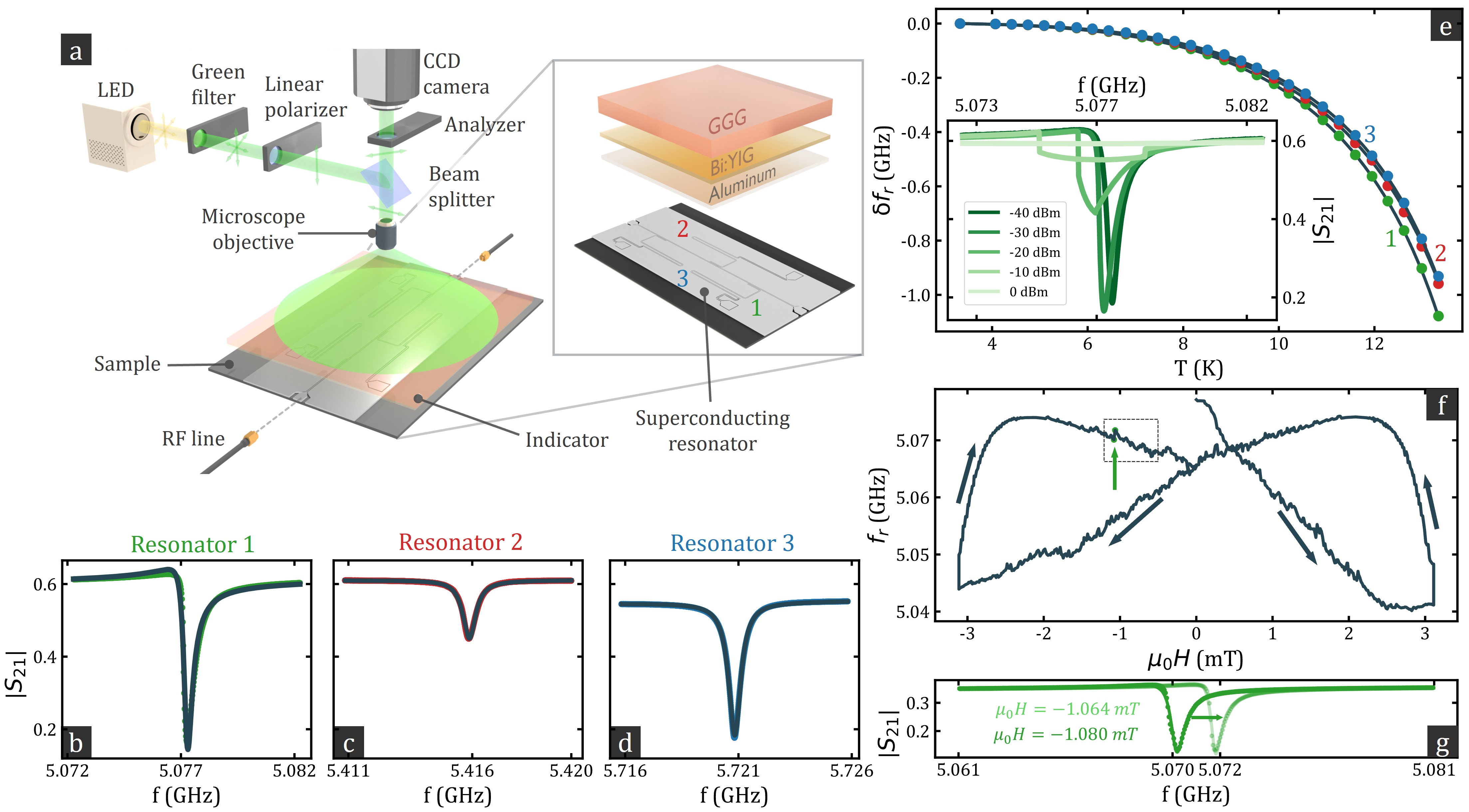}
\caption{(a) Experimental setup combining RF excitation of three multiplexed quarter-wavelength resonators and simultaneous magneto-optical imaging of the sample. The inset shows the typical components of the indicator placed on top of the device: the transparent gadolinium galium garnet (GGG) substrate, the 3 µm thick optically active indicator Bi:YIG, and a fully reflective mirror. In the final experiments, the mirror has been removed to avoid eddy currents induced in this metallic layer. In addition, the inset shows the geometry of the sample along with the labeling as resonator 1, 2, and 3. (b)-(d) Transmission coefficient $S_{21}$ in the vicinity of the three resonance frequencies at 3.3 K and with a RF power of -30 dBm. The black curves represent the fit obtained by following the procedure explained in \cite{khalil_analysis_2012}. (e) Relative change of the resonance frequency, $\delta f_r = f_r(T) - f_r(T = 3.33$ K),  as a function of temperature. The lines are fit to \autoref{Eq1}. The inset shows the resonance frequency peak shape as a function of excitation power. (f) Magnetic field dependence of the resonance frequency for an applied out-of-plane magnetic field, cycled from 0 mT $\rightarrow$ 3.3 mT $\rightarrow$ -3.3 mT $\rightarrow$ 0 mT, with dark arrows indicating the sweeping direction. The green arrow indicates the situation pictured in the panel (g) and corresponding to the transmission parameter for two consecutive applied magnetic fields. The dotted box indicates the magnetic field interval studied in \autoref{Fig4}.}\label{Fig1}
\end{figure*}

Two different layouts have been investigated. Device \#1, contains three  overcoupled $\lambda$/4 resonators covering a larger area and having a layout similar to the one used in Ref. \cite{nulens_non-invasive_2024} of lengths 4392 µm, 4561 µm, and 4744 µm, that are capacitively coupled to a central feedline with a width $w$ = 20 µm and a gap $g=10$ µm. Quality factors on the order of 10$^3$-10$^4$ are found for all bare resonators. Device \#2 has a surface of 2.5 $\times$ 2.5 mm$^2$ with a substrate size of 4 $\times$ 5 mm$^2$. Similar devices have been used in Ref. \cite{nulens_catastrophic_2023}. The samples contain two overcoupled, hanger-type $\lambda$/4 resonators of lengths 4089 µm and 3953 µm, that are capacitively coupled to a central feedline with a width $w$ = 20 µm and a gap $g=10$ µm. The resonators in sample \#1 are limited to a single U-turn and finished with an unused DC flux-biasing line as shown in the inset of \autoref{Fig1}(a), whereas for sample \#2 the resonators are meander-like with five U-turns. For the sake of clarity, we will present the results concerning sample \#1, and defer the results for sample \#2 to the Supplementary Material \cite{SM}.

\subsection{Magneto-optical and radio-frequency measurements}

The magnetic flux landscape was directly visualized by means of magneto-optical imaging (MOI). This technique exploits the Faraday rotation of linearly polarized light within a 3 µm-thick Bi-doped yttrium iron garnet indicator film with in-plane magnetic domains, placed in direct contact with the sample. As the polarization rotation scales with the local out-of-plane magnetic field component, the use of an analyzer oriented 4$^\circ$ off from perfect orthogonality to the incident polarization yields images whose intensity is proportional to this component and allows to distinguish between positive and negative fields. Image acquisition was performed using a CCD camera. Post-acquisition processing, carried out using a custom-made Python module, was employed to correct for non-uniform illumination and field-independent background contributions. \autoref{Fig1}(a) shows a simplified illustration of the MOI setup, while a detailed description is available in Ref. \cite{shaw_quantitative_2018}. Low-temperature measurements were conducted in a closed-cycle cryostat with a cooling power of approximately 100 mW at 4 K. The external magnetic field was applied by means of a copper coil with an experimentally determined sweep rate of 44 mT/s. MOI enables spatial mapping of the magnetic field, providing direct information on the magnetic flux distribution and the positions and extents of magnetic avalanches. Concomitant RF measurements were conducted using a Vector Network Analyzer (VNA). This instrument permits the direct acquisition of the forward transmission coefficient $S_{21}$, which quantifies the ratio of signal transmitted from Port 1 to Port 2. The power indicated throughout this work corresponds to the output power from the VNA. The actual power reaching the sample is about -10 dB below the VNA output.

\section{Results and discussion}

\subsection{Sample Characterization}

In order to characterize the response and performance of the planar resonators, we measured the $S_{21}$ transmission parameter as a function of frequency $f$, as shown in \autoref{Fig1}(b)-(d). The notch-filter response can be fitted following the procedure detailed in Ref. \cite{khalil_analysis_2012} as shown with a continuous curve. The three resonances observed around 5.08, 5.42, 5.72 GHz correspond to resonators 1, 2, and 3, respectively, as labeled in the inset of panel (a). The presence of bias lines leads to the distortion of the resonance peaks, indicating additional losses and the presence of parasitic modes introduced by the bias-line geometry \cite{nulens_non-invasive_2024}. In particular, resonator 1 exhibits a marked asymmetric shape due to overcoupling, whereas a lower quality factor is observed for resonator 2.


The resonance frequency of a quarter-wavelength resonator is determined by the expression $f_r=\left(4l\sqrt{(L_g'+L_k')C'}\right)^{-1}$, where $l$ corresponds to the resonator length, and $C'$, $L_g'$, and $L_k'$ are the capacitance, geometric inductance, and kinetic inductance per unit length, respectively. The relative change in resonance frequency $\delta f_r=f_r(T)-f_r(T_0)$, with $T_0=3.33$ K, as a function of temperature for the three resonators is shown in \autoref{Fig1}(e). This frequency shift results from the temperature-dependent kinetic inductance, and can be fitted using the following relation for conventional superconductors \cite{annunziata_tunable_2010},

\begin{equation}\label{Eq1}
    L'_k(T) = \frac{l}{w}\frac{hR_\square}{2\pi\Delta(T)}\frac{1}{\tanh\left( \frac{\Delta(T)}{2k_BT}\right)},
\end{equation}


\noindent where $w$ is the width of the resonator, $R_\square$ is the normal state sheet resistance, $h$ is the Planck constant, $k_B$ is the Boltzmann constant, and $\Delta(T)$ is the superconducting gap which can be expressed as \cite{tinkham_introduction_2004}:
\begin{equation}\label{Eq2}
    \Delta(T) = \Delta(0)\tanh\left( 1.74 \sqrt{\frac{T_c}{T}-1}\right),
\end{equation}

\noindent being $\Delta(0)$ the superconducting energy gap at zero temperature. Combining \autoref{Eq1} and \autoref{Eq2}, we obtain an expression of the temperature dependent resonance frequency. This can then be fitted, as pictured in \autoref{Fig1}(e). This procedure gives the following fitting parameters for resonator 1, resonator 2, and resonator 3, respectively: $L'_{g, 1} = (0.544 \pm 0.003)$ nH/mm, $C'_{1} = (0.161 \pm 0.008)$ nF/m, $L'_{k0, 1}=(0.1230 \pm 0.0006)$ nH/mm, $T_{c, 1} =( 14.51 \pm 0.45)$ K, $L'_{g, 2} = (0.562 \pm 0.003)$ nH/mm, $C'_{2} = (0.152 \pm 0.008)$  nF/m, $L'_{k0, 2}=(0.1080 \pm 0.0005)$ nH/mm, $T_{c, 2} =( 14.61 \pm 0.46)$ K and $L'_{g, 3} = (0.421 \pm 0.003)$ nH/mm, $C'_{3} = (0.202 \pm 0.002 $ nF/m, $L'_{k0, 3}=(0.0690 \pm 0.0005)$ nH/mm, $T_{c, 3} =( 14.56 \pm 0.62)$ K

The power dependence of the resonance dip presented in the inset of \autoref{Fig1}(e) reveals a gradual decrease in resonance frequency as the probing power increases. At RF powers exceeding -20 dBm, the transmission spectrum shows a steep, discontinuous jump in transmission on the low-frequency side of the resonance dip. Note that at 0 dBm the resonance peak is already entirely suppressed. The abrupt change in the shape of the resonance peak indicates that the resonator has entered a nonlinear regime, characterized by bifurcations in the $S_{21}$ transmission spectrum. This effect may arise from the power-dependent current in the resonator and the nonlinear kinetic inductance, as observed and described, for instance, in Ref. \cite{Swenson,Abdo,Ku}. Other mechanisms may develop at elevated microwave power (above 0 dBm), such as hot-spot formation \cite{Hampel,Kurter}. Increasing the RF power naturally leads to an increase in the RF magnetic field, which can eventually exceed the vortex penetration field (i.e., a few mT amplitude) and thereby produce microwave hysteresis losses. This effect in YBa$_2$Cu$_3$O$_{7-\delta}$ and NbN thin films for large RF fields were discussed in Ref. \cite{Nguyen}. Interestingly, the appearance of magnetic flux avalanches in MgB$_2$ \cite{ghigo_evidence_2007,ghigo-2009} and Pb \cite{awad_flux_2011,lara_thermally_2017} films has been reported in a large range of input power from -30 dBm to +5 dBm, even for RF magnetic fields smaller than the penetration field and frequencies smaller than the depinning frequency. These results suggest that local shaking of the vortex lattice (Campbell regime) may be enough to favor the triggering of magnetic flux avalanches. 


The resonance frequency of the planar resonators is also susceptible to the out-of-plane component of the applied field. Sweeping the out-of-plane magnetic field while tracking the resonance frequency reveals the characteristic butterfly-like shape \cite{bothner_magnetic_2012}, illustrated in \autoref{Fig1}(f). The critical state model for hard superconductors can naturally account for the observed irreversible response \cite{bothner_improving_2017}, through the magnetic field dependence of the surface impedance or the effective penetration of the RF field  \cite{Brandt-AC,Coffey,Pompeo}. Interestingly, a noisy response, in which the resonance frequency undergoes sudden jumps, is observed over an extensive range of applied magnetic fields. One of such resonance frequency jumps is shown in \autoref{Fig1}(g) and corresponds to the point indicated with an green arrow in panel (f). This stochastic response has also been reported for measurements under an in-plane magnetic field \cite{Li}, likely due to unwanted out-of-plane field components. It is worth mentioning that the actual magnetic field felt by the superconducting circuit can exceed, by an order of magnitude or more, the applied magnetic field due to the flux focusing effect \cite{Mawatari,bothner_improving_2017}. Nulens \textit{et al.} \cite{nulens_catastrophic_2023} amply discussed the different regimes encountered during out-of-plane magnetic field sweep, by combining {\it sequential} MOI and RF measurements. In this work, we aim to perform {\it simultaneous} RF transmission measurements and magnetic field imaging.


\subsection{Influence of the magnetic indicator}

\begin{figure*}[!ht]
\centering
\includegraphics*[width=1\linewidth]{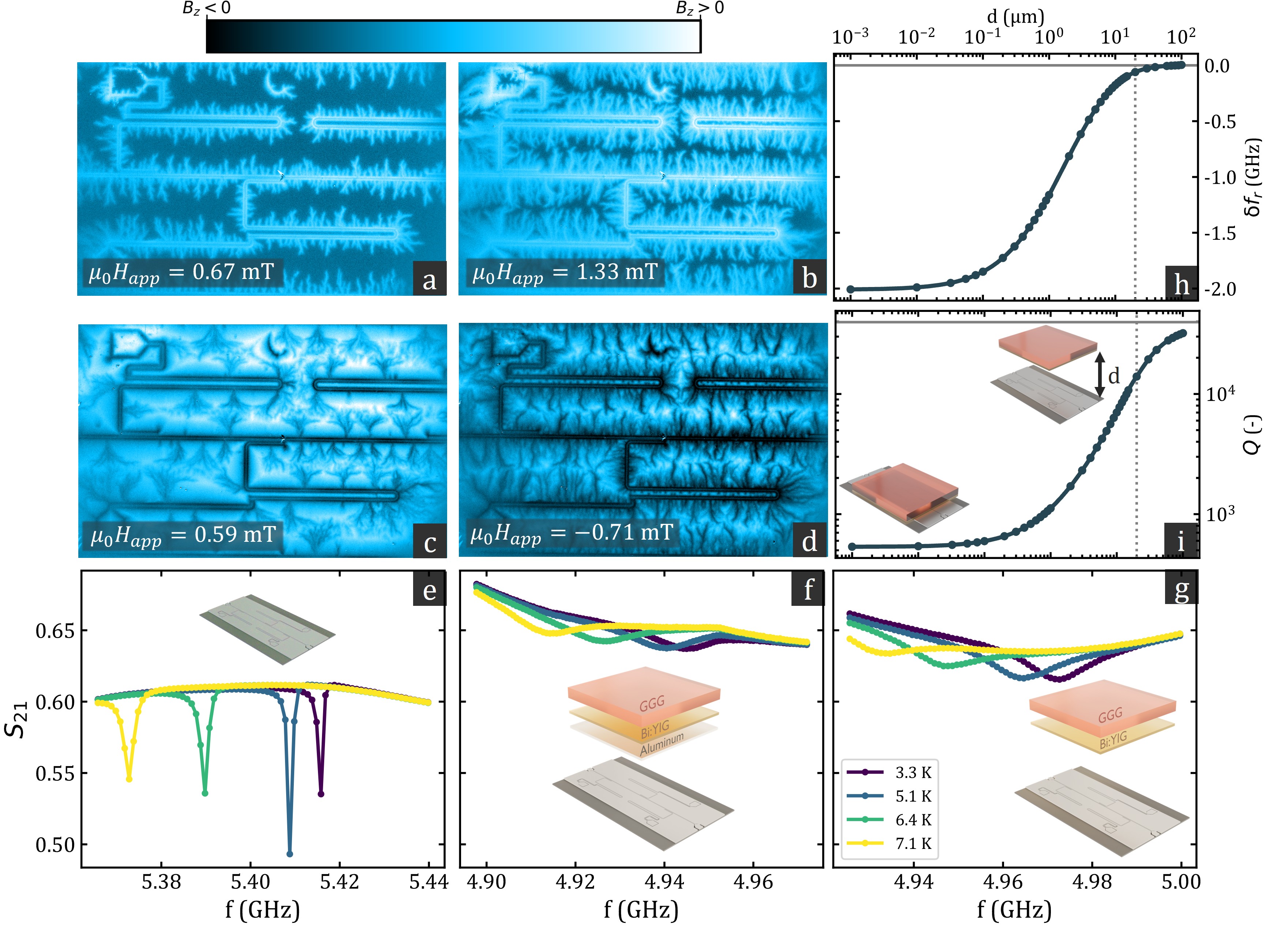}
\caption{(a)-(d) Magnetic flux penetration after a zero-field cooling during a magnetic field sweep from 0 mT $\rightarrow$ 3.3 mT $\rightarrow$ -3.3 mT $\rightarrow$ 0 mT at a base temperature of 3.2 K. Images were taken at 0.67 mT ($\mu_0 H_{app} \nearrow$), 1.33 mT ($\mu_0 H_{app} \nearrow$), 0.59 mT ($\mu_0 H_{app} \searrow$) and -0.71 mT ($\mu_0 H_{app} \searrow$), respectively. (e)-(g) Transmission coefficient $S_{21}$ for the resonator alone, with the indicator and with the indicator without mirror, respectively, at 3.3, 5.1, 6.4 and 7.1 K. (h, i) Evolution of the relative change in resonance frequency, $\delta f_r = f_r(d) - f_r(d = 100~\text{µm})$, and quality factor with decreasing distance between sample and indicator. The gray lines represent the width of the central microstrip line.}\label{Fig2}
\end{figure*}

To perform MOI on the planar devices, an optically active indicator is placed on top of the sample, as illustrated in \autoref{Fig1}(a). This allows one to capture the magnetic landscape during a magnetic field sweep, such as those displayed in \autoref{Fig2}(a)-(d). These images show the evolution of flux penetration and the proliferation of magnetic flux avalanches during the magnetic field sweep at $T=3.2$ K. These abrupt magnetic flux bursts develop from all the edges of the superconducting film, equally affecting the ground plane, the feedline, and the resonators. As reported in Ref. \cite{nulens_catastrophic_2023}, the flux avalanche activity in NbTiN develops at temperatures $T< 10$ K, substantially lower than the superconducting critical temperature. 

Unfortunately, the presence of the magnetic indicator significantly influences the transmission spectrum and the resonance parameters. This is illustrated in \autoref{Fig2}(e-g).
\autoref{Fig2}(e) shows the $S_{21}$ spectrum for different temperatures in the absence of a magnetic indicator. When the magnetic indicator is placed on top of the sample (\autoref{Fig2}(f)), the resonance peaks become broader, and the resonance frequency decreases. Here, the magnetic indicator has an aluminum bottom layer acting as a mirror. This conducting layer absorbs dissipative eddy currents induced by the RF field, thus severely affecting the performance of the resonators. A trade-off can be found by removing the metallic layer of the indicator, as shown in \autoref{Fig2}(g). In this case, the shift in resonance frequency persists, but the resonant lines are slightly better discerned than with the mirror. In addition, because the sample is reflective, little loss of MOI signal is observed, whereas the mixed topographic and magnetic signals can be separated via background postprocessing. Upon removing the mirror, seemingly delayed magnetic flux avalanches can be observed even up to a minute after increasing the magnetic field. The lack of this observation in the case of an indicator with mirror can be explained by the fact that the thin metal layer can suppress or deflect the propagation of flux avalanches through eddy currents generating opposing magnetic fields, known as magnetic braking \cite{Colauto,brisbois_classical_2014}. In addition, the enhanced thermal conductivity of the mirror layer might facilitate heat dissipation, thereby reducing the thermal feedback loop responsible for triggering additional flux avalanches.

The above analysis leads us to adopt the mirrorless indicator as the final configuration for carrying out simultaneous imaging and transmission measurements. This invasive character of the MOI technique on the RF performance of the resonators can even lead to a complete suppression of the resonance line, as illustrated in the Supplementary Material for sample \#2 \cite{SM}. 

To further understand the effect of placing the indicator without mirror on the device, we resort to numerical simulations using the commercial software Sonnet \cite{noauthor_sonnet_nodate} for which we took into account the exact dimensions of the bilayer indicator with $\varepsilon=11.99$ for the GGG substrate \cite{connelly_complex_2021} and $\varepsilon=10$ for the Bi:YIG \cite{sirdeshmukh_dielectric_1998}. \autoref{Fig2}(h) and (i) show the relative change in resonance frequency and quality factor as a function of the distance between sample and indicator. The results show a significant reduction in resonance frequency and quality factor with decreasing distance. Notably, unlike the resonance frequency, the quality factor exhibits a reduction even at relatively large distances, where the shift in resonance frequency remains relatively weak. It can be seen that the resonance frequency only becomes significantly affected at distances of the order of the central track line width (20 µm), as highlighted by the gray dotted line in \autoref{Fig2}(h,i). Furthermore, both resonance parameters reach a plateau at small distances, indicating that the influence of the indicator saturates when it is brought into close proximity with the resonator. The experimentally observed shift of less than 0.5 GHz, together with numerical simulations, indicates a separation of approximately 5 µm between the resonators and the indicator. This value is consistent with typical distances between sample and indicator as reported in Ref. \cite{shaw_quantitative_2018}. 

This numerical study highlighted how the resonator is affected by the proximity of the indicator, which is consistent with our experimental findings of the resonance frequency shift and quality factor reduction. It is worth mentioning that, the observed resonance shift is not easily reproducible across successive installations of the indicator, likely due to variations in its positioning and distance from the sample surface. The observed downshift can be attributed to the increase in the effective permittivity of the medium surrounding the coplanar waveguide, resulting from the presence of the indicator. Since $f_r \propto 1/\sqrt{\varepsilon_r}$ \cite{goppl_coplanar_2008}, a decrease in resonance frequency is expected. Furthermore, as the resonator is affected by the proximity of the indicator, dielectric losses increase and, consequently, a lower quality factor can be observed.

These results highlight the critical influence of the spacing between the sample and the indicator. On the one hand, as the indicator is brought closer to the sample, its invasiveness increases. On the other hand, achieving high MOI resolutions requires minimizing the distance between the sample and the indicator to reduce the spread of magnetic flux lines away from the sample. This trade-off underscores the experimental challenge of performing MOI while concurrently monitoring the resonance frequency.

\subsection{Effect of RF excitation on the magnetic avalanche activity}
\begin{figure*}[!ht]
\centering
\includegraphics*[width=1\linewidth]{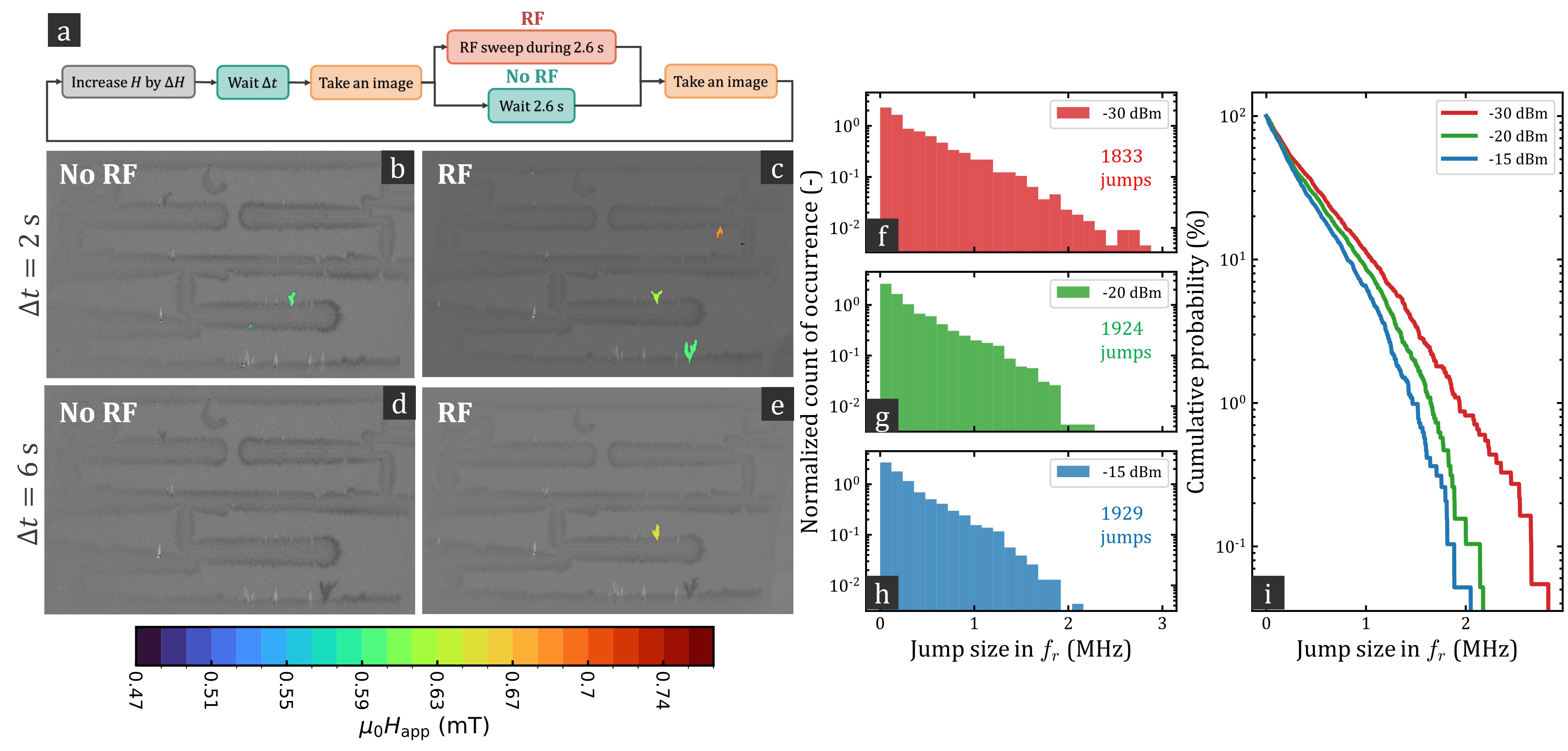}
\caption{(a) Schematic of the experimental protocol followed to characterize the influence a RF sweep on magnetic flux avalanches. This protocol is applied after a ZFC. (b)-(e) Avalanche activity appearing after waiting $\Delta t$ when increasing the magnetic field. (b) and (c) show the avalanches appearing after $\Delta t =2$ s for a control measurement and during a RF sweep respectively. (d) and (e) show the avalanches appearing after $\Delta t=6$ s for the control measurement and during the RF sweep respectively. (f)-(h) Histogram of the frequency jump sizes appearing during a magnetic field sweep for -30 dBm, -20 dBm and -15 dBm. Each graph results from 10 magnetic field sweeps with a constant field step of 0.0216 mT. (i) Cumulative probability of the frequency jump sizes for the three RF powers applied.}\label{Fig3}
\end{figure*}

Next, we investigate the effect of the RF excitation on the nucleation and propagation of the magnetic flux avalanches. To that end, the magnetic field is swept in a stepwise fashion, following a complete loop (0 $\rightarrow$ 3.2 mT $\rightarrow$ -3.2 mT $\rightarrow$ 0) and, at each magnetic-field step, a frequency sweep around the corresponding resonance frequency is performed. Before and after this RF sweep, we take magneto-optical images in order to visualize the avalanches appearing during the RF excitation. To determine whether the observed avalanches are RF-induced or simply of delayed nature, we perform a control measurement in which we do not sweep the frequency but instead wait the same amount of time before taking a second image. This experimental protocol is summarized in \autoref{Fig3}(a). \autoref{Fig3}(b)-(e) show representative MOI images displaying the avalanches appearing during the RF sweep (labeled "RF"), or in the waiting time for the control measurement (labeled "no-RF"). The individual avalanches are colored with respect to the field value at which they were detected. \autoref{Fig3}(b) and (c) are captured in the case in which the waiting time after increasing the magnetic field, $\Delta t$, was 2 s, whereas \autoref{Fig3}(d) and (e) result from a waiting time of $\Delta t =$ 6 s. For larger delay times, the avalanche activity is too weak and hiinders meaningful statistical analysis. It can be noted that the number of avalanches is reduced when increasing the delay time $\Delta t$, hinting at the fact that some of these avalanches' nature is a delayed one instead of an RF-induced avalanche. However, when comparing with the control measurement, it can be seen that there are significantly more avalanches in the RF case.

In order to investigate the effect of increasing RF power on the triggering of flux avalanches, we analyze the $f_r(H)$ response and track the amplitude of the resonance shift caused by the abrupt flux penetration. \autoref{Fig3}(f)-(h) show the distribution of the sizes of the jumps in resonance frequency during a field sweep after zero-field cooling. It can be noted that the tail of the histogram extends further into large jump sizes at lower RF powers. Additionally, when comparing the number of jumps detected, higher RF powers result in slightly more jumps of the resonance frequency. This indicates that at lower RF power, bigger jumps in resonance frequency occur, although they are less frequent. \autoref{Fig3}(i) shows the same experimental data in the form of the cumulative distribution of the frequency jump sizes.  This cumulative distribution corresponds to the integral of the probability distribution, which translates to the probability of having a jump bigger or equal than the abscissa. While the cumulative probability curves for the three powers overlap for small step sizes, they depart from each other for large frequency shifts. Higher RF powers result in steeper curves, indicating that larger jumps are less likely.

To summarize, the finding presented above indicate that vortex shaking by the RF field $\mu_0 h_{RF}$ within the low-power regime ($\mu_0 h_{RF} < H_p$, where $H_p$ denotes the vortex penetration field) slightly enhances magnetic flux avalanche activity, but it seems to remain a marginal effect, negligible when comparing the impact of changing the applied DC field. It is interesting to point out that we have not found any evidence of more avalanche activity at the position where the circulating current (or the RF field ) maximizes. This is consistent with previous reports suggesting that it is not needed for $\mu_0 h_{RF}$ to exceed the vortex penetration field, in order to trigger avalanches. Concerning the observation that at higher RF powers, large jumps occur less often, one should note that avalanches are the result of a metastable state that decays into a closer-to-equilibrium state. The more gentle the shaking is, the further away from equilibrium this metastable is, and therefore the larger the resulting avalanche.

\subsection{Effect of the magnetic activity on the RF transmission}

In the previous section, we discussed the influence of the RF excitation on the thermomagnetic instabilities. Let us now analyse the opposite case concerning the impact of avalanche activity on the resonance frequency. The simultaneous RF-MOI measurements enable us to extend the investigation beyond the previous report by Nulens et al. \cite{nulens_catastrophic_2023}. Indeed, in Ref. \cite{nulens_catastrophic_2023}, it was demonstrated that a noisy $f_r(H)$ response can be associated with the presence of magnetic flux avalanches. However, it was not possible to unambiguously identify which individual flux avalanche was the source of a given resonance frequency jump. The concomitant RF-MOI measurements permit to bridge this gap.  

\begin{figure*}[!ht]
\centering
\includegraphics*[width=1\linewidth]{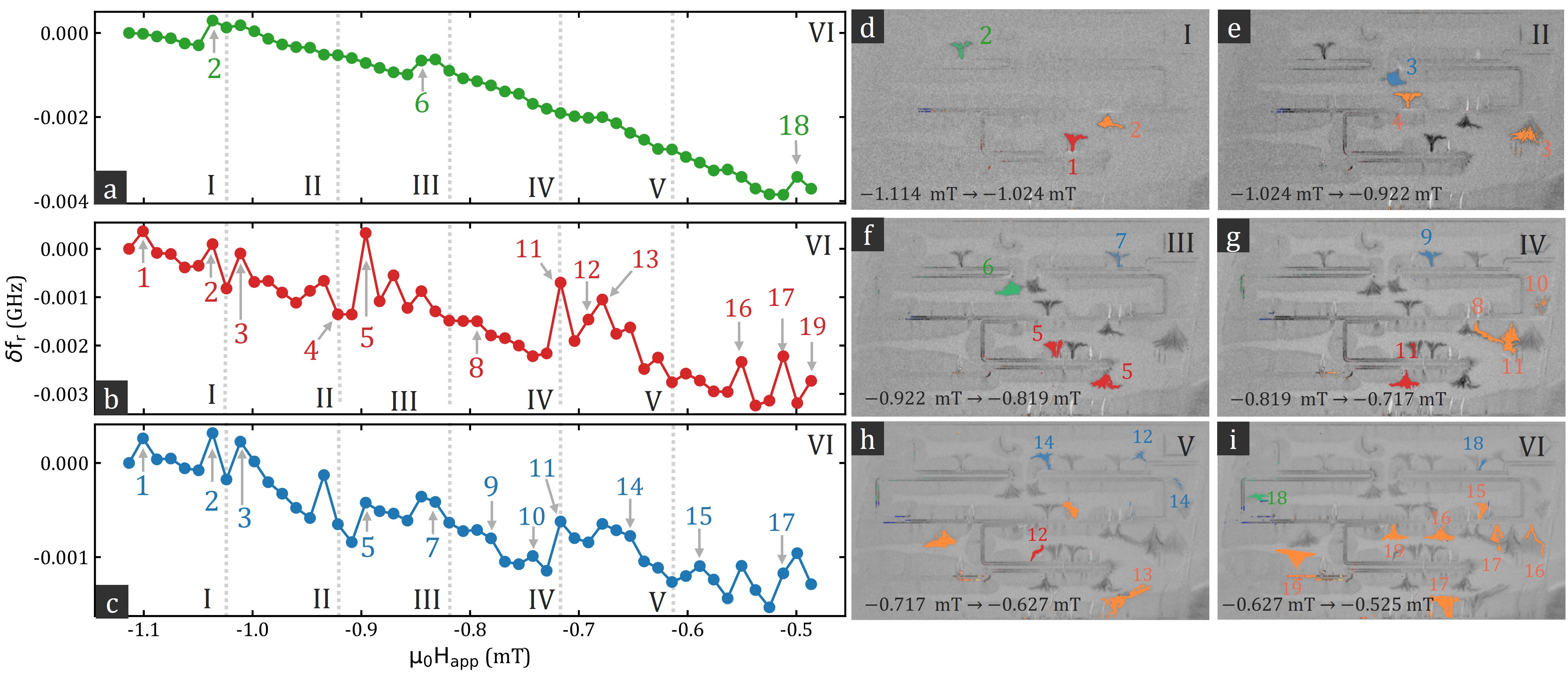}
\caption{(a)-(c) Resonance frequency as a function of applied magnetic field of resonator 1, 2 and 3, respectively. (d)-(i) MOI images related to the intervals of magnetic field defined in (a)-(c). Each new avalanche is colored with respect to where it nucleates, where resonator 1 is green, resonator 2 is red, resonator 3 is blue, whereas avalanches forming at the feedline and the border edges are colored in orange. The field range and magnetic history correspond to the dotted box indicated in Figure 1(f).}\label{Fig4}
\end{figure*}

\autoref{Fig4}(a-c) shows the resonance frequency for the three resonators on the chip during a magnetic field sweep along one of the branches of the $f_r(H)$ loop, as indicated with a dotted box in \autoref{Fig1}(f). Representative snapshots showing the magnetic activity during the magnetic field intervals, and marked with Roman numerals, are shown in the left panels. A complete animation covering a wider magnetic field range is available in the Supplementary Material \cite{SM}. In the snapshots, each emerging avalanche is assigned a color code that permits identification of the nucleation sites. Avalanches nucleating in resonator 1 are colored green, on resonator 2 with red, and on resonator 3 with blue. Avalanches forming anywhere else, specifically at the feedline or the sample's edges, are colored in orange. Each avalanche event has been labeled with a number and linked to the corresponding jump in resonance frequency in panels (a-c).

It can be seen that jumps in resonance frequency correlate with magnetic flux avalanches nucleating on the corresponding resonator. Upward jumps can be explained by an effective decrease of the magnetic field at the edges of the patterned sample after an avalanche event. This is so because of the particular magnetic history in which the sample is initially fully penetrated by a positive magnetic field, and avalanches of anti-vortices tend to decrease the local average magnetic field. Since the RF current distribution is inhomogeneous and peaks at the edges \cite{Lahl2005}, a decrease in the local magnetic field also results in a reduction of dissipation caused by vortex shaking. Additionally, a decrease in the magnetic field reduces the kinetic inductance, thereby increasing the resonance frequency. After this event and until the next avalanche occurs, the resonance frequency is observed to decrease as magnetic pressure builds up again. Downward jumps are also observed, and they are actually more frequent and cover a larger magnetic field range than upward jumps. Their origin can be understood considering the fact that an avalanche effectively creates a normal region with a local increase in magnetic flux, and therefore with higher kinetic inductance. 



Interestingly, we found that events in one resonator may affect the neighboring resonators. This non-local effect is manifested, for instance, in event 1, which is triggered in resonator 2, and affects the resonance frequency not only of resonator 2, but also that of resonator 3. Similarly, event 2, which involves two simultaneous avalanches, one located in resonator 1 and another along the excitation line, this event is detected in all three resonators. This cause-effect non-locality is intimately related to the fact that, for thin film geometry in a perpendicular magnetic field, the magnetic field at a given position is determined by the distribution of the current density over the whole film \cite{Brandt-Indembom}. The immediate consequence of this observation is that strategies for improving flux pinning need to consider the non-locality of the problem. In other words, reinforcing pinning there where the density of vortices maximizes may only be a partial solution to a mitigation plan. 

It may also come as a surprise to see that magnetic flux avalanches are favored along flat borders, rather than corners, where the concerted action of current crowding and nanofabrication defects is expected to facilitate the penetration of vortices. This counterintuitive observation occurs at intermediate temperatures and has been explained in Ref. \cite{Jiang-2020,silhanek-review} based on the field dependence of the critical current. Nevertheless, at low temperatures, we have observed that avalanches first appear at the U-shaped turn of the resonators.

One cannot help but note that in resonator 1, avalanches are markedly less frequent than in the other two resonators. We attribute this observation to the fact that this resonator is effectively at a higher temperature. The technical reason is that the size of the chip is slightly larger than the cold finger of the cryostat, making thermalization of this part of the substrate slightly less efficient. Since avalanche activity decreases with increasing temperature, this resonator is expected to exhibit fewer avalanche events. In the Supplementary Material \cite{SM}, we show a similar analysis to that presented in \autoref{Fig4} for other field ranges.



\begin{figure*}[!ht]
\centering
\includegraphics*[width=1\linewidth]{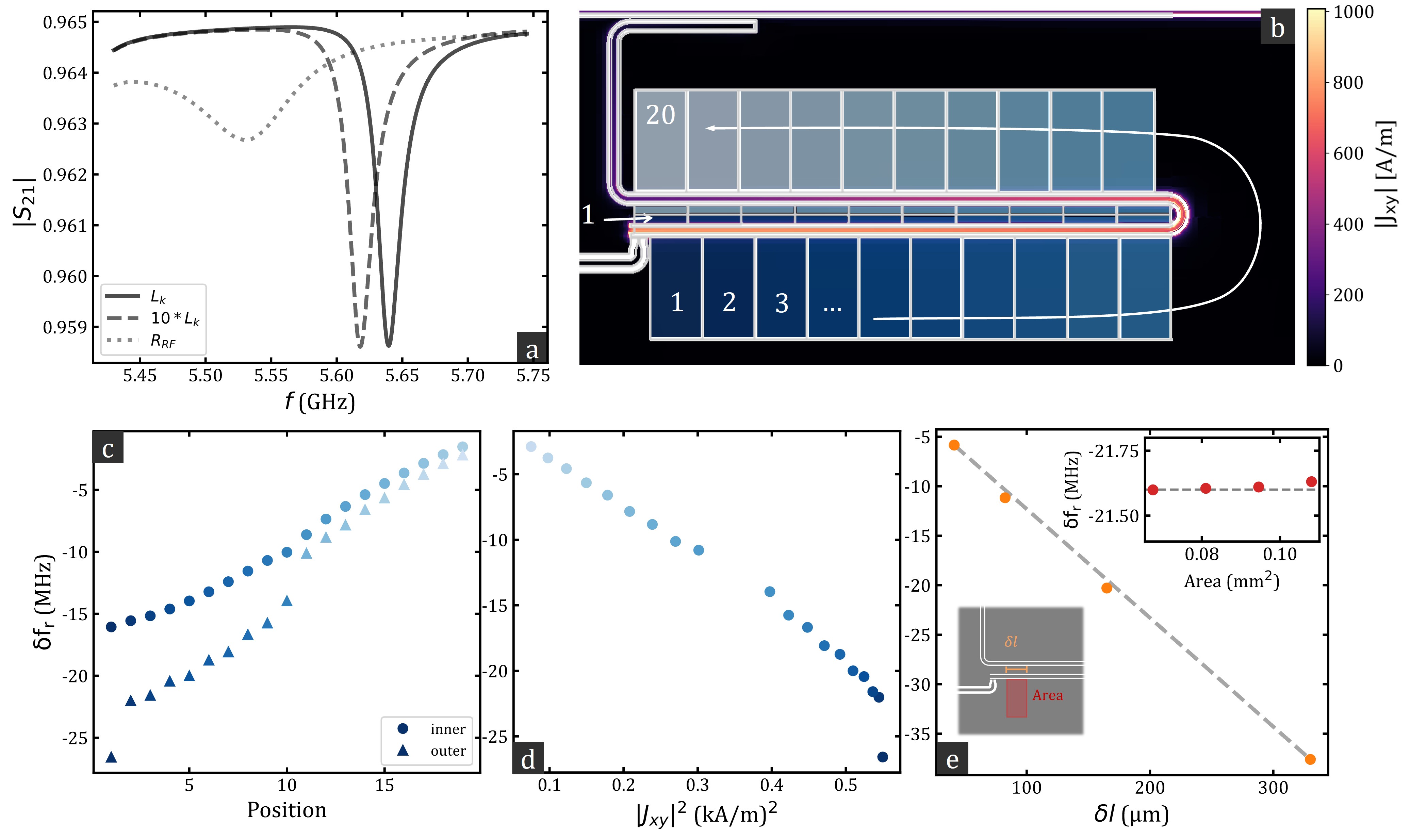}
\caption{(a) Simulated $S_{21}$ spectrum of the resonance peak for a superconducting area that is modeled using the same kinetic inductance ($L_k$) as the rest of the material, an increased kinetic inductance (10*$L_k$) and finally, with a resistive component ($R_{RF})$. (b) Illustration of the simulated current density at resonance at an applied power of +13 dBm, superposed with the subdivisions considered in the following. (c) Simulated relative change in resonance frequency as a function of the position of the simulated area. The triangle markers corresponds to the outer areas and the dots to the inner ones, as shown in (b). Furthermore, the marker colors correspond to the position of the considered areas.
(d) Change in resonance frequency as a function of the current density squared at the center conductor near the simulated area. (e) Simulated relative change in resonance frequency as a function of the contact length with the edge of the conductor. The upper inset shows the change in resonance frequency as a function of the area of the simulated patch, while the lower inset illustrates the contact length and area of the simulated patch.}\label{Fig5}
\end{figure*}

The origin of the observed variability in the size of the frequency jump can be manifold: (i) the multiplicity of events co-occurring, (ii) the size of the avalanche, and (iii) the physical location of the avalanche. Unfortunately, it is not straightforward to disentangle the contribution of each of these factors based on the experimental observations. Instead, we resort to the method of moment simulations of the electromagnetic response of the resonators using Sonnet, in which mimicking a magnetic flux avalanche can be achieved either by a local increase in resistance or a local rise in kinetic inductance. As shown in Fig. \ref{Fig5}(a), the former approach leads to a decrease of $f_r$ and simultaneously to a reduction of $Q$. In contrast to that, a local decrease of $L_k$ produces a reduction of $f_r$ but with little change of $Q$. When comparing with the experimental result presented in Fig. \ref{Fig1}(g), for which no net reduction of $Q$ is observed, one can naturally propose to emulate the avalanches by a local change of $L_k$. This approximation can be physically understood by considering that an avalanche renders a region of the superconductor resistive, but only for a short time. At the moment the $S_{21}(f)$ measurement is carried out, the avalanche region has already cooled down below the superconducting critical temperature, and can be considered as a region with a {\it frozen} high local magnetic field. This high local field leads to an increase of $L_k$ and therefore to a reduction of the resonance frequency. As explained above, in the case of a fully penetrated sample in which avalanches of anti-vortices occur, the average magnetic field in the avalanche trunk decreases, and therefore $L_k$ decreases, thus leading to an increase of the resonance frequency. 

Figure \ref{Fig5}(c-e) summarizes the main findings of this numerical investigation for the geometrical layout presented in panel (b), exactly matching the geometry of the measured sample. In these simulations, the superconductor is considered a perfect conductor, and avalanches are represented by rectangular regions with a kinetic inductance an order of magnitude larger than that of the unaffected areas. By selecting the location of the avalanches along the resonator (see panel (b)), we can estimate the impact on the resonance frequency (panel (c)). Figure \ref{Fig5}(b) shows 20 possible avalanches of identical size along the external perimeter of the resonator, and another 20 avalanches along the internal perimeter of the U-turn. In the same panel, the amplitude of the sheet current density is indicated for an input power of +13 dBm. Figure \ref{Fig5}(c) shows the calculated drop in resonance frequency $\delta f_r$ as a function of position along the resonator. Several observations can be drawn from these simulations. 

First, flux avalanches have a higher impact on $f_r$ where the amplitude of the sheet current density is higher. This correlation is made more apparent in panel (d), where $\delta f_r$ of the 20 possible avalanche locations is plotted as a function of the amplitude of the sheet current density squared. Previous studies have similarly shown that shifts in the resonant frequency can be traced back to the surface magnetic field and, by extension, to the surface current density, thereby confirming that the measured frequency shifts provide insight into the amplitude of the current distribution \cite{Culbertson,Lejeune-2025}.

Second, avalanches appearing at the inner part of the U-shaped meander seem to have a lower impact than avalanches at the outer perimeter. This could be attributed to the fact that inner avalanches cover a substantially smaller surface than outer ones. However, we have verified that only the contact length matters (see panel (e)), not the surface of the avalanche (see inset of panel (e)). Thus, one is tempted to conclude that avalanches at the inner border of the resonator are less harmful than those at the outer border. However, care must be exercised with this interpretation, since the simulations incorporate the effect of the magnetic field through the local changes in $L_k$, which need to be adapted accordingly for the inner and outer rims of the resonator. Indeed, the field-focusing effect at the inner border should lead to a proportionally larger $L_k$ for inner avalanches compared to outer ones. 

Third, we observe that avalanches occurring at one resonator do not influence the response of the neighboring resonators. This lack of non-locality is expected since the non-linear electrodynamic characteristic of superconducting thin films is not accounted for in the model. Furthermore, we do not observe an influence of avalanches at the feedline on the resonator.

As a concluding discussion, we aim to clarify the dynamic regimes explored in this study and delineate the range of parameters covered. First, it is important to acknowledge that the presence of vortices in superconducting resonators is inevitable. Most resonators are designed to operate in the gigahertz range, which typically requires a fabrication scale of several millimeters. Considering the case of NbTiN with superconducting characteristic lengths $\xi \sim 4$ nm and $\lambda \sim 400$ nm, there is plenty of space to host a large amount of vortices \cite{Nowack}. Since RF currents are localized along the edges, vortices located in the perimeter of the structure are particularly deleterious. Concerning avalanches, a much larger length-scale $\ell^*$ dominates the underlying physics. This length represents the threshold of magnetic flux penetration depth above which thermomagnetic instabilities may develop \cite{denisov_onset_2006, denisov_dendritic_2006}. In the samples examined in this study, avalanches were clearly observed within the 60‑µm-wide spacing between the two branches of the U-shaped resonators. Additionally, the images suggest localized flux penetration in both the central superconducting line of the resonators and the feedline. 

It has been recently shown \cite{Nowack} that in a 20-µm-wide NbTiN strip under field-cooling conditions, vortex trapping can occur at magnetic fields as low as 10 µT. Considering that the lower penetration field is $B_{c1} \approx 4$ mT, and taking into account the field magnification due to demagnetization effects $\sim \sqrt{W/t}$, with $W=5$ mm the sample's width, and $t$ its thickness, we estimate a first penetration field into the outer rim of the sample of $B_p \sim 20$ µT. In this estimation we are assuming that Bean-Livingston barriers are suppressed by borders defects and inhomogeneities \cite{silhanek-review}. The first discontinuity in the resonance frequency is observed at much higher fields (100 - 500 µT), and corresponds to the magnetic perforation of the outer rim and sudden injection of magnetic flux into the gaps of the structure, as described in Ref. \cite{nulens_catastrophic_2023}. This magnetic field is then locally increased by a factor of $\sim 10$ due to field focusing effect, as explained in Ref. \cite{Mawatari,bothner_improving_2017}, and rapidly populates with vortices the borders of feedline, resonator, and ground plane.
Concerning the RF excitation, the presented measurements span a power range between -30 dBm and 0 dBm which corresponds to circulating current densities $J_{RF} \sim 0.1-3 \cdot 10^5$ A/cm$^2$, and a RF fields $\mu_0 h_{RF} \sim 0.03-1$ mT (after considering the field focusing effect \cite{Mawatari}). These currents and fields are respectively smaller than the critical current density of NbTiN ($J_{c} \sim 1 \cdot 10^7$ A/cm$^2$ \cite{Nowack}) and the lower critical field, thus suggesting that the radio-frequency excitation is limited to vortex shaking rather than reconfiguration of the vortex distribution. Assuming a depinning frequency $f_{dp} \sim 10$ GHz \cite{Janju}, we can conclude that the measurements in this work have been performed within the linear Campbell regime \cite{Pompeo}. This is consistent with the threshold $\mu_0 h_{RF} \sim 5$ mT for non-linear response for NbN films \cite{Nguyen}. It is worth pointing out that the magnetic flux avalanches develop in the ground plane or in the spacing between the two branches of the U-shaped resonators, where even lower current densities are expected \cite{Lahl2005}. Indeed, the edge currents in the ground plane, next to the excitation line or the resonators are about a factor $1+2g/w \sim 2$ smaller than in the central line \cite{Culbertson}.

\section{Conclusions}

In summary, simultaneous radio-frequency transmission measurements and magnetic imaging enabled the investigation of the influence of RF excitation on magnetic flux avalanche activity, and, conversely, shed light on the abrupt changes in resonance frequency caused by the flux avalanches. Although clear evidence suggests that avalanche activity increases when the radio-frequency field is on, the impact of the RF in the linear Campbell regime is relatively minor. We demonstrated that the MOI technique used to image the magnetic flux is invasive, resulting in a downshift of the resonance frequency and a reduction in the quality factor. This influence was further analyzed using the method of moment simulations of the MOI indicator approaching the device. Alternatively, less intrusive techniques, such as Quantum Diamond Microscopy \cite{shulga2025}, could be implemented to make the reduction of the threshold field for triggering magnetic flux avalanches under RF excitation more apparent. We were able to tag every jump in resonance frequency to a particular avalanche event, thereby demonstrating the non-locality of the electrodynamics: resonant frequency jumps can result from events occurring away from the resonator. This is relevant information for optimizing the resonator's design in terms of out-of-plane magnetic field tolerance. Method of moment simulations further clarify the impact of the magnetic flux avalanches according to their nucleation point along the resonator. It is worth pointing out that the magnetic flux avalanche regime is ubiquitous in thin films made of conventional superconductors and therefore may affect most of the technologically mature quantum devices. In contrast, high-$T_c$ superconducting thin films are less susceptible to these events and may offer an alternative route to magnetic-field-resilient resonators.

\begin{acknowledgments} 
\subsection*{Acknowledgments}
The authors acknowledge financial support from Fonds de la Recherche Scientifique - FNRS under the grant Weave PDR T.0208.23, and the Research Foundation and by COST (European Cooperation in Science and Technology) [www.cost.eu] through COST Action SUPERQUMAP (CA 21144). N. L. acknowledges support from FRS-FNRS (Research Fellowships FRIA).
\end{acknowledgments}

\bibliography{references}

\end{document}